\documentclass[onecolumn,aps,prd,superscriptaddress,preprintnumbers,nofootinbib,10pt]{revtex4-2}
\usepackage{amsfonts}
\usepackage[tbtags]{amsmath}
\usepackage{amssymb}
\usepackage{color}
\usepackage{courier}
\usepackage{dsfont}
\usepackage{euscript}
\usepackage{epsfig}
\usepackage{epstopdf}
\usepackage{float}
\usepackage{graphics}
\usepackage{graphicx}
\usepackage{hyperref}
\usepackage[utf8]{inputenc}
\usepackage{latexsym}
\usepackage{MnSymbol}
\usepackage{pifont}
\usepackage{physics}
\usepackage{slashed}
\usepackage{subfigure}
\usepackage{tikz-feynman}
\usepackage[normalem]{ulem}
\usepackage{url}
\usepackage{verbatim}
\usepackage{widetable}
\usepackage{setspace}

\allowdisplaybreaks

\hypersetup{
	colorlinks=true,
	citecolor=blue,
	citebordercolor=red,
	linktoc=all,
	linkcolor=blue,
	urlcolor=blue
}

\newcommand{\1}{\mathbf{1}}

\newbox{\ORCIDicon}
\sbox{\ORCIDicon}{\large\includegraphics[width=0.8em]{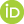}}

\begin{document}

	\title{{\bf \Large  Bell and Mermin inequalities in Quantum Field Theory \\from vacuum projectors and Weyl operators }}
	
	\vspace{1cm}

	\author{M. S.  Guimaraes}\email{msguimaraes@uerj.br} \affiliation{UERJ $–$ Universidade do Estado do Rio de Janeiro,	Instituto de Física $–$ Departamento de Física Teórica $–$ Rua São Francisco Xavier 524, 20550-013, Maracanã, Rio de Janeiro, Brazil}
	
	\author{I. Roditi} \email{roditi@cbpf.br} \affiliation{CBPF $-$ Centro Brasileiro de Pesquisas Físicas, Rua Dr. Xavier Sigaud 150, 22290-180, Rio de Janeiro, Brazil }
	
	
	\author{S. P. Sorella} \email{silvio.sorella@fis.uerj.br} \affiliation{UERJ $–$ Universidade do Estado do Rio de Janeiro,	Instituto de Física $–$ Departamento de Física Teórica $–$ Rua São Francisco Xavier 524, 20550-013, Maracanã, Rio de Janeiro, Brazil}

	\author{A. F. Vieira\,\href{https://orcid.org/0000-0003-2897-2437}{\usebox{\ORCIDicon}}} \email{arthurfvieira@if.ufrj.br} \affiliation{UFRJ $-$ Universidade Federal do Rio de Janeiro, Instituto de Física, RJ 21.941-972, Brazil}

	\vspace{2cm}

	\begin{abstract}
		The use of the vacuum projector $|0 \rangle \langle 0| $ and of the unitary Weyl operators enables us to construct a set of Hermitian dichotomic operators in relativistic scalar Quantum Field Theory in Minkowski spacetime. Employing test functions supported in diamond regions, both Bell and Mermin inequalities are studied by means of a numerical setup. In addition to reporting expressive violations of both inequalities, the cluster property is also checked. 
		
	\end{abstract}
	
	\maketitle

	\section{Introduction}\label{s1}
	
	It seems safe to state that the study of the Bell-CHSH inequality \cite{Bell:1964kc,Clauser:1969ny} in relativistic Quantum Field Theory\footnote{See  \cite{Guimaraes:2024mmp} for a recent review on the subject.} has not yet reached the same level of comprehension and of concrete calculability which has been achieved in Quantum Mechanics. Despite the remarkable theorems established by \cite{Llandau,Summers:1987fn,Summers:1987squ,Summers:1987ze} within the context of Algebraic Quantum Field Theory \cite{Haag:1992hx}, several aspects remain to be unraveled.  For instance, we might quote:
	\begin{itemize} 
		
		\item construction of a suitable set of Hermitian dichotomic field operators, as required by the Bell-CHSH inequality, see \cite{Guimaraes:2024byw};
		
		\item specification of the compact supported causal test functions associated to the Minkowski regions selected for the analysis of the    inequality;
		
		\item formulation of a computational setup able to  explicitly implement the results proven in \cite{Summers:1987fn,Summers:1987squ,Summers:1987ze}  for the vacuum state of the theory and for causal regions like: complementary wedges and double  tangent diamonds; 
		
		\item check of other relevant relationships containing the physical parameters of the theory. This is the case of the cluster property, a key feature of Quantum Field Theory \cite{Haag:1992hx}, which exhibits an explicit dependence from the mass parameter as well as from the spatial distance between the two spacelike regions for which the Bell-CHSH inequality is being tested. 
		
	\end{itemize}
	
	In this work, we shall elaborate on all these issues, with the aim of establishing a computational numerical setup to detect the violations of both Bell-CHSH and Mermin \cite{Mermin} inequalities in diamond-shaped regions in $1+1$ Minkowski spacetime, building upon our previous attempts  \cite{Dudal:2023mij,DeFabritiis:2024jfy,Guimaraes:2024xtj}.
	
	More precisely, in Sect.\eqref{vp}, we shall present the construction of an explicit set of Hermitian dichotomic operators for a real scalar quantum field. The procedure relies on the use of the vacuum projector and Weyl unitary operators. In Sect.\eqref{B},  the violation of the Bell-CHSH inequality will be established in the case of two causal tangent diamonds. Here, we present a discussion of the low-mass limit, which turns out to be in very nice agreement with the results of \cite{Summers:1987fn,Summers:1987squ,Summers:1987ze}. Sect.\eqref{M3} will be devoted to the violation of the Mermin inequality of order three \cite{Mermin}. Further, in Sect.\eqref{Cp}, we provide a numerical check of the cluster property for the correlation functions of the Bell operators. Section \eqref{conc} presents our conclusions, while Appendix \eqref{appA} provides a brief overview of the canonical quantization of the massive real scalar field.

	\section{Vacuum projector and construction of Hermitian dichotomic field operators}\label{vp}
	
\begin{figure}[t!]
		\begin{minipage}[b]{0.6\linewidth}
			\includegraphics[width=\textwidth]{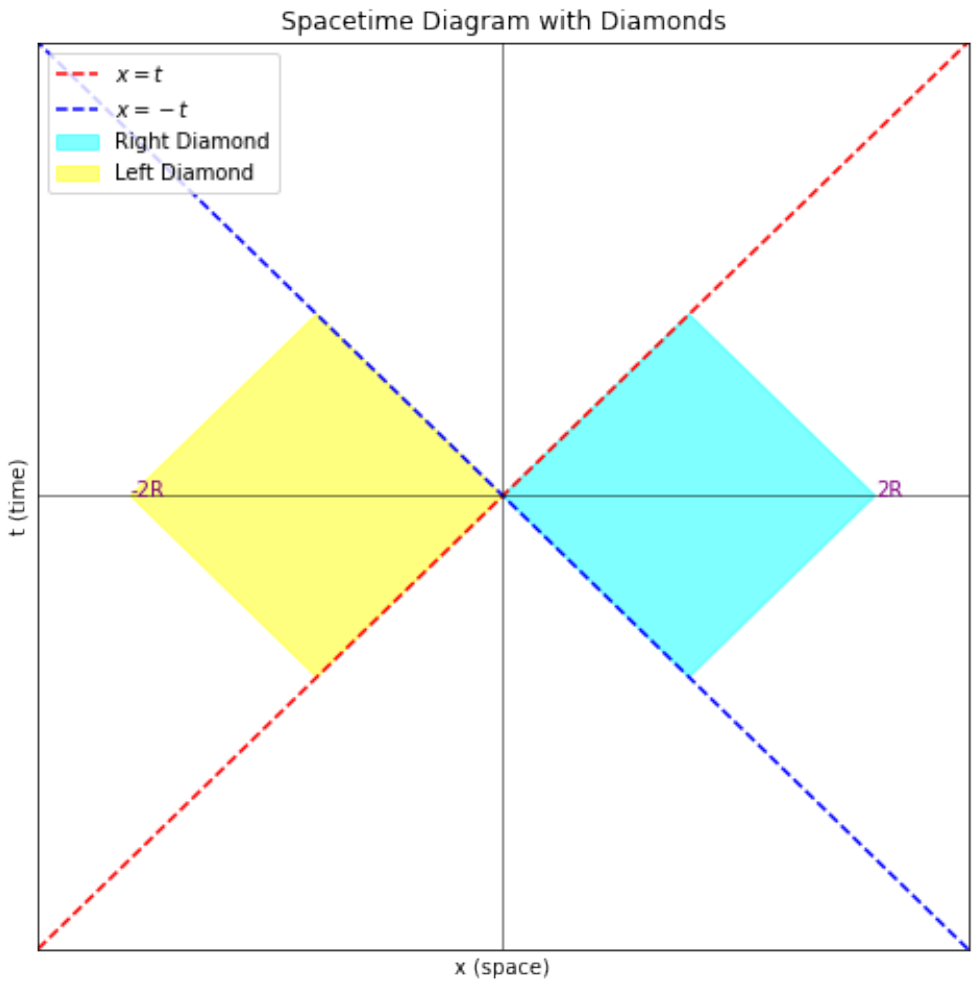}
		\end{minipage} \hfill
		\caption{Causal tangent diamond regions.  The right diamond corresponds to the region $\{ (x,t), \; |x-R| + |t| \le R \}$. A similar equation holds for the left diamond. The origin is the contact point between the two diamonds. }
		\label{diamonds}
	\end{figure}

	In order to construct suitable operators to capture the violation of the Bell-CHSH inequality, we start by considering the vacuum projector $\ket{0}\bra{0}$,
	where $|0\rangle$ is the vacuum state defined by the condition \cite{Haag:1992hx}
	\begin{equation} 
		a_h |0\rangle = 0, \; \; \forall h \;. \label{vst}
	\end{equation} 
	Here, $a_h$ stands for the smeared annihilation operator (see Appendix \eqref{appA}) 
	\begin{equation} 
		a_h = \int \frac{dk}{2\pi} \frac{1}{2 \omega_k} h^{*}(\omega_k,k) a_k \;, \qquad  a_h^{\dagger} = \int \frac{dk}{2\pi} \frac{1}{2 \omega_k} h(\omega_k,k) a_k^{\dagger}  \;, \qquad\label{smah}
	\end{equation}
	with 
	\begin{equation} 
		[a_h, a_{h'}^{\dagger}] = \langle h | h'\rangle = \int \frac{dk}{2\pi} \frac{1}{2 \omega_k} h^{*}(\omega_k,k) h'(\omega_k,k) = \frac{i}{2}\Delta_{PJ}(h,h') + H(h,h') \;, \label{innp}
	\end{equation}
	where $\langle h|h'\rangle$ is the Lorentz invariant inner product, Eq.~\eqref{InnerProduct}, and  $h(\omega_k,k)$ denotes the Fourier transform of a smooth test function $h(t,x)$ with compact support. The expressions $\Delta_{PJ}(h,h')$ and $H(h,h')$ denote the smeared Pauli-Jordan and Hadamard distributions, respectively, Eq.~\eqref{mint}.\\\\As mentioned previously, we shall consider two tangent causal diamond regions, as depicted in Fig.\eqref{diamonds}.
	The right diamond will be referred to as Alice's diamond, while the left one will be referred to as Bob's diamond. Let us denote by $(f,f')$ the pair of Alice's test functions: two smooth functions supported in right diamonds. Similarly, $(g,g')$ will denote Bob's test functions, {\it i.e.,} two smooth functions supported in left diamonds. Of course, the supports of $(f,f')$ are spacelike with respect to those of $(g,g')$:
	\begin{equation} 
		supp(f,f')  \;\;\; {\rm spacelike} \;\;\; supp(g,g') \;. \label{supp}
	\end{equation} 
	Out of the vacuum projector, we introduce the dichotomic operator 
	\begin{equation} 
		{\cal F} = \mathbf{1} -2 \;|0 \rangle \langle 0| \;, \qquad {\cal F}^2=\1 \;,\label{dicF}
	\end{equation} 
	as well as the Weyl unitaries \cite{Summers:1987fn,Summers:1987squ,Summers:1987ze,DeFabritiis:2023tkh,Guimaraes:2024mmp,Azevedo:2024zeb}:
	\begin{equation} 
		W_h = e^{i \varphi(h)} \;, \qquad W_h^{\dagger} W_h= W_h W_h^{\dagger} =\1 \;, \qquad W^{\dagger}_h = W_{(-h)} \;, \label{Weyl}
	\end{equation}
	where $ \varphi(h) $ is the smeared field,  Eq.~\eqref{smmd}. The Weyl operators obey the following relation 
	\begin{equation} 
		W_h \; W_{h'}= e^{-\frac{i}{2} \Delta_{PJ}(h,h')}\; W_{(h+h')} \;, \label{Walg}
	\end{equation}
	with $\Delta_{PJ}(h,h')$ being the Pauli-Jordan smeared distribution, Eq.~\eqref{mint}. Therefore, following \cite{Guimaraes:2024mmp,Azevedo:2024zeb}, Alices's and Bob's operators are obtained upon acting with the Weyl unitaries on the operator ${\cal F}$, namely
	\begin{eqnarray} 
		A_f & = & W^{\dagger}_f \; {\cal F} \; W_f = e^{-i \varphi(f)}\; (\1 -2 \;|0 \rangle \langle 0|)\;e^{i \varphi(f)} \;, \qquad A_{f '}  =  W^{\dagger}_{f'} \; {\cal F} \; W_{f'} = e^{-i \varphi(f')}\; (\1 -2 \;|0 \rangle \langle 0|)\;e^{i \varphi(f')} \;, \nonumber \\
		B_g & = & W_g \; {\cal F} \; W^{\dagger}_g = e^{i \varphi(g)}\; (\1 -2 \;|0 \rangle \langle 0|)\;e^{-i \varphi(g)} \;, \qquad B_{g '}  =  W_{g'} \; {\cal F} \; W^{\dagger}_{g'} = e^{i \varphi(g')}\; (\1 -2 \;|0 \rangle \langle 0|)\;e^{-i \varphi(g')} \;. \label{ABop}
	\end{eqnarray}
	These operators fulfill the conditions to be admissible \cite{Llandau,Summers:1987fn,Summers:1987squ,Summers:1987ze} for the study of the Bell-CHSH ineqiality:
	\begin{eqnarray} 
		A_f & = & A^{\dagger}_f \;, \qquad A_f^2 =\1 \;, \qquad A_{f'} = A_{f'}^{\dagger} \;, \qquad A_{f'}^2 = \1 \;, \nonumber \\
		B_g & = & B^{\dagger}_g \;, \qquad B_g^2 =\1 \;, \qquad B_{g'} = B_{g'}^{\dagger} \;, \qquad B_{g'}^2 = \1 \;, \nonumber \\
		&& [A_f, A_{f'}]  \neq  0 \;, \qquad [B_g, B_{g'}] \neq 0 \;. \label{Bcds}
	\end{eqnarray}
Furthermore, with regard to the commutation relations between $(A_f,A_{f'})$ and $(B_g,B_{g'})$, it can be readily verified that
\begin{equation} 
\langle 0\; [A_f ,B_g]\; |0\rangle  = \langle 0|\; [A_f,B_{g'}] \; |0\rangle=\langle 0|\; [A_{f'},B_{g}]\;|0\rangle =\langle 0|\; [A_{f'},B_{g'}] \;|0\rangle= 0 \;, \label{chc}
\end{equation}
so that $(A_f,A_{f'})$ and $(B_g,B_{g'})$ are useful operators for the study of the Bell-CHSH inequality in the vacuum state $|0\rangle$.

	\section{The Bell-CHSH inequality for causal tangent diamonds}\label{B} 
	
	After constructing a suitable set of operators, we can now proceed with the study of the Bell-CHSH inequality in the vacuum state.\cite{Summers:1987fn,Summers:1987squ,Summers:1987ze,DeFabritiis:2023tkh,Guimaraes:2024mmp,Azevedo:2024zeb}:
	\begin{equation} 
		\langle 0|\; {\cal C}\; |0\rangle = \langle 0|\; (A_f + A_{f'})B_g + (A_f -A_{f'})B_{g'} \; |0\rangle \;. \label{BCHSH}
	\end{equation} 
	A violation of the Bell-CHSH inequality occurs whenever 
	\begin{equation} 
		2 < | \langle 0|\; {\cal C}\; |0\rangle | \le 2 \sqrt{2} \;, \label{vbchsh}
	\end{equation}
	where the value $2\sqrt{2}$ is known as the Tsirelson bound \cite{Cirelson:1980ry}. Recalling that, for two spacelike supported test functions $(f,g)$, it holds that \cite{Summers:1987fn,Summers:1987squ,Summers:1987ze,DeFabritiis:2023tkh,Guimaraes:2024mmp,Azevedo:2024zeb}
	\begin{equation} 
		\langle 0 | \;e^{i \varphi(f)} e^{i \varphi(g)}\; |0\rangle = \langle 0 | \;e^{i (\varphi(f)+ \varphi(g)}\; |0\rangle = e^{-\frac{1}{2}||f+g||^2} = e^{-\frac{1}{2}(H(f,f)+H(g,g) + 2H(f,g))} \;, \label{Had}
	\end{equation} 
	where $ ||f+g||^2= \langle f+g|f+g\rangle$, it follows that 
	\begin {equation} 
	\langle 0|\; A_f B_g \; |0\rangle = 1 + 4 \;e^{-(H(f,f) + H(g,g) + H(f,g))} - 2 \;e^{-H(f,f)} - 2 \;e^{-H(g,g)} \;. \label{abvac}
\end{equation}
Therefore, for the Bell-CHSH correlator, one gets 
\begin{eqnarray} 
	\langle 0|\; {\cal C}\; |0\rangle  & = &  2 + 4 \;e^{-(H(f,f) + H(g,g) + H(f,g))} + 4 \;e^{-(H(f',f') + H(g,g) + H(f',g))} +4 \;e^{-(H(f,f) + H(g',g') + H(f,g'))} \nonumber \\
	& - & 4 \;e^{-(H(f',f') + H(g',g') + H(f',g'))}  - 4  \;e^{-H(f,f)} - 4  \;e^{-H(g,g)} \;. \label{chshcorr}
\end{eqnarray}

\subsection{The choice of the test functions}\label{tf}

\begin{figure}[t!]
\begin{minipage}[b]{0.4\linewidth}
\includegraphics[width=\textwidth]{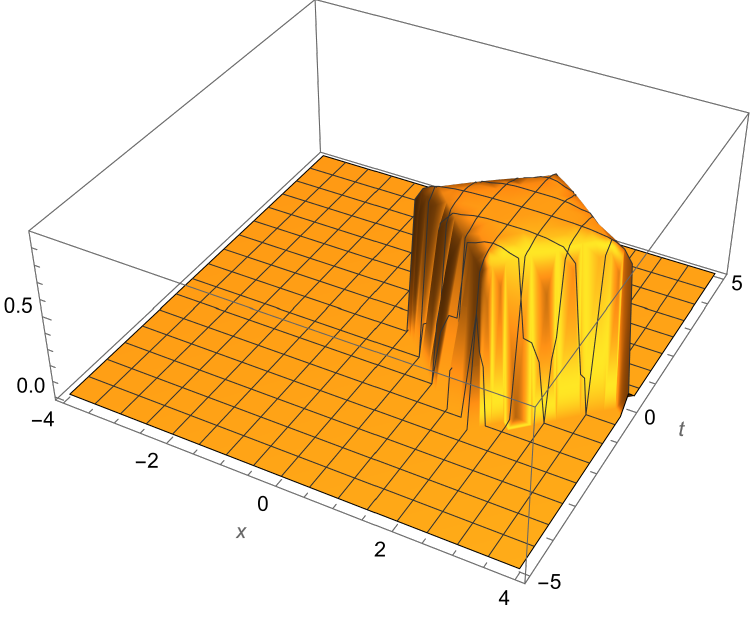}
\end{minipage} \hfill
\caption{Plot of the test function $f(t,x)$, for $(a=0.1, \eta=1, R=2 )$.   }
\label{fig.ff}
\end{figure}

\begin{figure}[t!]
\begin{minipage}[b]{0.4\linewidth}
\includegraphics[width=\textwidth]{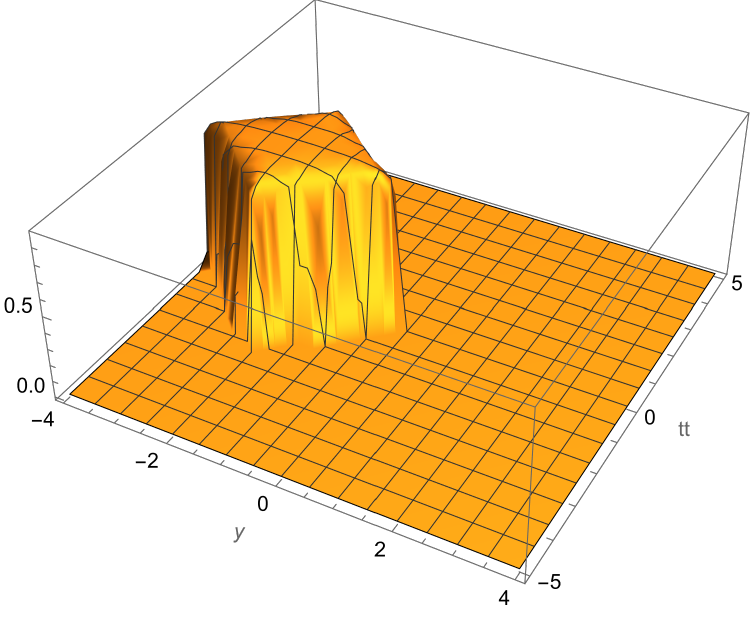}
\end{minipage} \hfill
\caption{Plot of the test function $g(t,x)$, for $(b=0.1, \sigma=1, R=2 )$.   }
\label{fig.gg}
\end{figure}

At this stage. we have to specify the shape of the test functions which will be employed. Let us consider first the right diamond, specified by the condition 
\begin{equation} 
	|x-R| + |t| \le R \;. \label{rd}
\end{equation} 
For $(f,f')$ we write 
\begin{align}
	f(t,x) = \eta
	\left\{
	\begin {aligned}
	& e^{-\frac{a}{R^2 - (|x-R| +|t|)^2}}, \quad & |x-R| + |t| \le R,   \\
	& 0 \quad  {\rm elsewhere}                   
\end{aligned}
\right. \label{ff}
\end{align}
and 
\begin{align}
f'(t,x) = \eta'
\left\{
\begin {aligned}
& e^{-\frac{a'}{R'^2 - (|x-R'| +|t|)^2}}, \quad & |x-R'| + |t| \le R',   \\
& 0 \quad  {\rm elsewhere}                   
\end{aligned}
\right. \label{ffp}
\end{align}
where $(a,a',R,R',\eta,\eta')$ are arbitrary parameters, to be fixed at the best convenience. \\\\Analogously, in the left diamonds, one  considers 
\begin{align}
g(t,x) = \sigma
\left\{
\begin {aligned}
& e^{-\frac{b}{R^2 - (|x+R| +|t|)^2}}, \quad & |x+R| + |t| \le R,   \\
& 0 \quad  {\rm elsewhere}                   
\end{aligned}
\right. \label{gg}
\end{align}
and 
\begin{align}
g'(t,x) = \sigma'
\left\{
\begin {aligned}
& e^{-\frac{b'}{R'^2 - (|x+R'| +|t|)^2}}, \quad & |x+R'| + |t| \le R',   \\
& 0 \quad  {\rm elsewhere}                   
\end{aligned}
\right. \label{ggp}
\end{align}
with $(b,b',\sigma,\sigma')$ free parameters. The behavior of $f$ and $g$ is shown in Figs.~\eqref{fig.ff} and \eqref{fig.gg}. One sees that $f$ vanishes in the left wedge, while $g$ vanishes in the right wedge. 

\subsection{The numerical setup}  

Let us now describe the numerical setup we adopted. Regarding the numerical integration of the scalar products in Eq.~\eqref{chshcorr}, the typical integral is shown in Eq.~\eqref{mint}. Due to the challenges in evaluating the Fourier transformation of the test functions in momentum space analytically, expression \eqref{mint} has been computed directly in configuration space. We used Mathematica, employing two integration methods: QuasiMonteCarlo and MultidimensionalRule. The parameters $(a,\eta, a', \alpha',\eta')$, $(b,\sigma, b',\sigma')$, along with the mass $m$ and $(R,R')$, were selected through tests using a random algorithm. For each test, $10^5$ random values for the parameters were evaluated. Table (\ref{tabelaMax}) presents an overview of the results obtained.





\begin{table}[h!]
\centering
\resizebox{\textwidth}{!}{\begin{tabular}[t]{|c|c|c|c|c|c|c|c|c|c|c|c| }
\hline
$a$ & $\eta$ & $b$ & $\sigma$ & $a'$ & $\eta'$ & $b'$ & $\sigma'$ &	$m$  & $R$ & $R'$ & 
$\langle \mathcal{C} \rangle$  \\
\hline 
0.0571763 & 0.173707 & 0.682824 & 0.0240641 &3.60771 & 0.784553 & 0.300806 & 1.70987 & 0.300647 &
501998 & 0.799741 &
2.029125 
\\	
\hline
0.710532 & 0.285758 & 0.248215 & 0.0876402 & 
0.472765 & 2.89372 & 3.65721 & 3.08397 & 0.000588745 & 
707315 & 0.710241 & 
2.0660 
\\
\hline 
0.753259 & 0.249479 & 0.413562 & 0.0140057 & 4.97831 & 4.43684 & 0.898361 & 7.15717 & 
4.14395 $\times 10^{-6}$ & 0.815919 & 0.752558 & 
2.093229
\\
\hline 
0.495696& 0.180809 & 0.471991& 0.087649 & 4.0448  & 4.4751 & 1.9839 &11.1014 & 2.62258 $ \times10^{-8}$ & 0.869138 &
0.867249 &  2.206017  
\\
\hline
\end{tabular}}
\caption{Results obtained for the Bell-CHSH correlation function \eqref{chshcorr}. The values of the violation are reported in the last column.}
\label{tabelaMax}
\end{table}	

\noindent One sees that the test functions \eqref{ff}-\eqref{ggp} give rise to significant violations of the Bell-CHSH inequality.

\subsection{Interpolation for low values of the mass parameter}\label{lm}

The general results obtained in \cite{Summers:1987fn,Summers:1987squ,Summers:1987ze} apply to the diamond configuration that we are analyzing here, see Fig.~\eqref{diamonds}. According to \cite{Summers:1987fn,Summers:1987squ,Summers:1987ze}, it holds that the Bell-CHSH inequality in the vacuum state and for double causal tangent diamonds achieves maximal violation for massless fields, {\it i.e.}, $2\sqrt{2}$. \\\\In the present case, as we are considering a scalar field in $1+1$ Minkowski spacetime, we cannot take the massless limit\footnote{Notice that, in $1+1$ spacetime, the Lorentz invariant integration measure collapses to an infrared divergent quantity, {\it i.e.} 
\begin{equation} 
\frac{dk}{2\pi} \frac{1}{2\sqrt{k^2+m^2}} \;\; \rightarrow\;\; \frac{dk}{2\pi} \frac{1}{2|k|} \;. \label{irm}
\end{equation}
}, due to the existence of infrared singularities. Despite this, we can gradually decrease the mass parameter and verify whether the outputs align with the previous statement. To assist the reader, we once again present the values of the Bell-CHSH violation alongside the corresponding mass, as shown in Table~\eqref{mass}.

\begin{table}[h!]
\begin{tabular}[t]{|c|c|}
\hline
$m$  &
$\langle \mathcal{C} \rangle$  \\
\hline 
0.0093905 & 2.06704 \\
\hline 			

0.000588745	& 2.0660\\	
\hline
4.14395 $\times 10^{-6}$ & 2.093229 \\
\hline 			
2.62258 $\times 10^{-8}$  & 2.206017 \\
\hline	

\end{tabular}
\caption{Violations of the Bell-CHSH correlator $\langle {\cal C}\rangle$ together with the corresponding values of the mass parameter $m$.  }
\label{mass}
\end{table}

 The parameter values $(a, \eta, b, \sigma, a', \eta', b', \sigma', R,R')$ corresponding to $\langle {\cal C}\rangle 
=(2.0660, 2.093229, 2.206017)$ are those already listed in Table \eqref{tabelaMax}. For the first value, $\langle {\cal C}\rangle = 2.06704$, the corresponding parameters are provided in Table \eqref{tabelaMob}.

\begin{table}[h!]
\centering
\resizebox{\textwidth}{!}{\begin{tabular}[t]{|c|c|c|c|c|c|c|c|c|c|c|c| }
\hline
$a$ & $\eta$ & $b$ & $\sigma$ & $a'$ & $\eta'$ & $b'$ & $\sigma'$ &	$m$  & $R$ & $R'$ & 
$\langle \mathcal{C} \rangle$  \\
\hline 
 0.453107
 & 0.06256 
 & 0.241230 
 & 0.033623 
 & 3.008120 
 & 4.486029 
 & 0.699209 
 &  4.096952 
 & 0.00939 
& 1.859616
& 0.840575 
&2.06704
\\
\hline
\end{tabular}}
\caption{Values of the paramters $(a,\eta,b,\sigma, a',\eta'.b', \sigma', R,R')$ corresponding to $\langle {\cal C} \rangle = 2.06704$.}
\label{tabelaMob}
\end{table}

\begin{figure}[t!]
\begin{minipage}[b]{0.5\linewidth}
\includegraphics[width=\textwidth]{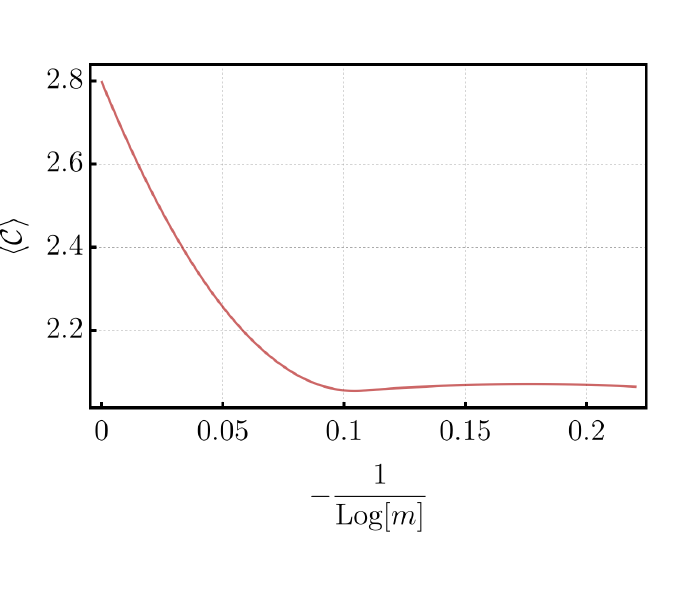}
\end{minipage} 
\caption{Interpolating curve obtained from the values of Table \eqref{mass}, showing the behaviour of the Bell-CHSH correlator 
$ \langle {\cal C} \rangle$ as function of the inverse logarithm of the mass parameter $m$. }
\label{int}
\end{figure}

One clearly sees that the size of the violation increases as the mass gets smaller, as expected from \cite{Summers:1987fn,Summers:1987squ,Summers:1987ze}. The values reported in Table \eqref{mass} can also be used to get an interpolating curve, reported in Fig.~\eqref{int} 

The qualitative agreement with \cite{Summers:1987fn,Summers:1987squ,Summers:1987ze} looks pretty good. The maximum value of the Bell-CHSH inequality, corresponding to $m=0$, is $\expval{\cal C} = 2.79824$, quite close to Tsirelson's bound of $2\sqrt{2}$. 
 We recall that a rigorous proof demonstrating that the Bell-CHSH inequality reaches its maximal value, {\it i.e.} $2 \sqrt{2}$,  in the case of double causal tangent diamonds (see Fig.\eqref{diamonds}) can be found in  \cite{Summers:1987fn,Summers:1987squ}.

\section{The Mermin inequality of order three}\label{M3}

\begin{figure}[t!]
\begin{minipage}[b]{0.4\linewidth}
\includegraphics[width=\textwidth]{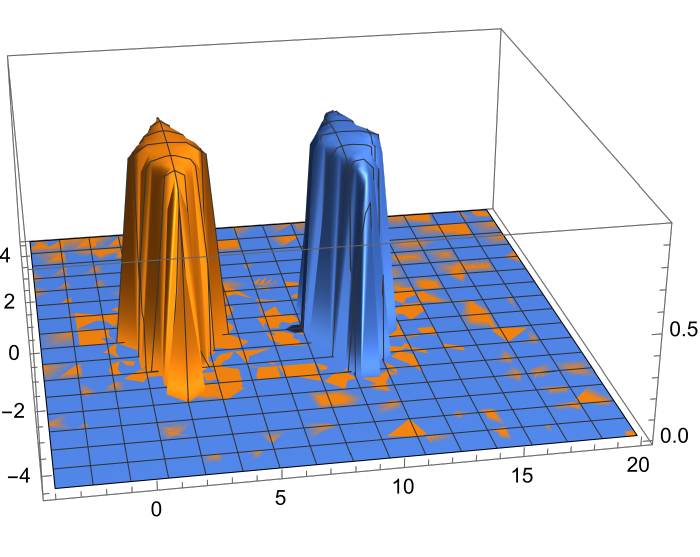}
\end{minipage} \hfill
\caption{Plot of the test functions $f(t,x)$ and $h(t,x)$:  $x\in [-4,20]$ and $t\in -4..5,4.5]$. The function $f(t,x)$ corresponds to the orange curve, while $h(t,x)$ to the blue one. The values of the parameters are as follows: $(a=0.1,\eta=1, R=2)$ and $(p=0.1, \zeta=1,d=2)$. }
\label{mmm4}
\end{figure}

While we have not yet achieved results for the Mermin inequalities \cite{Mermin} analogous to those established for the Bell-CHSH inequality \cite{Summers:1987fn,Summers:1987squ,Summers:1987ze}, the diamond regions allow us to adapt the previous numerical setup to this context. As a concrete example, we will consider the Mermin inequality of order three:
\begin{equation} 
\langle 0|\; {\cal M}_3 \; |0\rangle = \langle 0|\; A_{f'} B_g C_h + A_f  B_{g'} C_h + A_f B_g C_{h'} - A_{f'} B_{g'} C_{h'}  \; |0\rangle \;, \label{mthree}
\end{equation} 
where $(A_f,A_{f'}, B_g, B_{g'}, C_h,C_{h'})$ are Hermitian dichotomic operators subject to the conditions 
\begin{eqnarray} 
A_f & = & A^{\dagger}_f \;, \qquad A_f^2 =\1 \;, \qquad A_{f'} = A_{f'}^{\dagger} \;, \qquad A_{f'}^2 = \1 \;, \nonumber \\
B_g & = & B^{\dagger}_g \;, \qquad B_g^2 =\1 \;, \qquad B_{g'} = B_{g'}^{\dagger} \;, \qquad B_{g'}^2 = \1 \;, \nonumber \\
C_h & = & C^{\dagger}_h \;, \qquad C_h^{2}=\1 \;, \qquad C_{h'} = C_{h'}^{\dagger} \;, \qquad C^2_{h'} =\1 \;, \nonumber \\
&&\langle 0| [A_f ,B_g] |0 \rangle = \langle 0|[A_f,B_{g'}] |0\rangle =0 \;, \qquad \langle 0| [A_{f'},B_{g}] |0\rangle =\langle 0| [A_{f'},B_{g'}]|0\rangle = 0 \;, \nonumber \\
&& \langle 0| [A_f,C_h] |0 \rangle = \langle 0|[A_{f},C_{h'}]|0\rangle =0 \;, \qquad \langle 0| [A_{f'},C_h] |0\rangle =\langle 0| [A_{f'},C_{h'}]|0\rangle =0 \;, \nonumber \\
&&\langle 0| [B_g,C_h] |0\rangle = \langle 0| [B_{g},C_{h'}] |0\rangle =0 \;, \qquad \langle 0| [B_{g'},C_h] |0\rangle = \langle 0| [B_{g'},C_{h'}]|0\rangle =0 \;, \nonumber \\
&& [A_f, A_{f'}]  \neq  0 \;, \qquad [B_g, B_{g'}] \neq 0 \;, \qquad [C_h,C_{h'}] \neq 0 \;, \label{mcds}
\end{eqnarray}
where, similarly to the operators $(A,B)$ of Eqs.~\eqref{ABop},
\begin{equation} 
C_h  = W^{\dagger}_h \; {\cal F} \; W_h = e^{-i \varphi(h)}\; (\1 -2 \;|0 \rangle \langle 0|)\;e^{i \varphi(h)} \;, \qquad C_{h '}  =  W^{\dagger}_{h'} \; {\cal F} \; W_{h'} = e^{-i \varphi(h')}\; (\1 -2 \;|0 \rangle \langle 0|)\;e^{i \varphi(h')} \;. \label{cop} 
\end{equation}
A violation of the Mermin inequality takes place whenever \cite{Mermin}
\begin{equation} 
2 < | \langle 0|\; {\cal M}_3 \; |0\rangle | \le 4 \;. \label{vm}
\end{equation}
The new pair of test functions $(h,h')$ is demanded to be spacelike supported with respect to $(f,f')$ and $(g,g')$:
\begin{align}
h(t,x) = \zeta
\left\{
\begin {aligned}
& e^{-\frac{p}{R^2 - (|x-d-3R| +|t|)^2}}, \quad & |x-d-3R| + |t| \le R,  \; \quad x\in[d+2R,\, d+4R],  \\
& 0 \quad  {\rm elsewhere}                   
\end{aligned}
\right. \label{hh}
\end{align}
and 
\begin{align}
h'(t,x) = \zeta'
\left\{
\begin {aligned}
& e^{-\frac{p'}{R'^2 - (|x-d'-5-3R'| +|t|)^2}}, \quad & |x-d'-3R'| + |t| \le R', \; \quad x\in[d'+2R',\, d'+4R'].   \\
& 0 \quad  {\rm elsewhere}                   
\end{aligned}
\right. \label{hhp}
\end{align}
Here, $(p,p',\zeta,\zeta')$ and $(d,d')$ are free parameters that can be selected to satisfy the aforementioned spacelike condition. This goal is accomplished by using three diamond regions, as illustrated in Fig.~\eqref{mmm}:

\begin{figure}[t!]
\begin{minipage}[b]{0.7\linewidth}
\includegraphics[width=\textwidth]{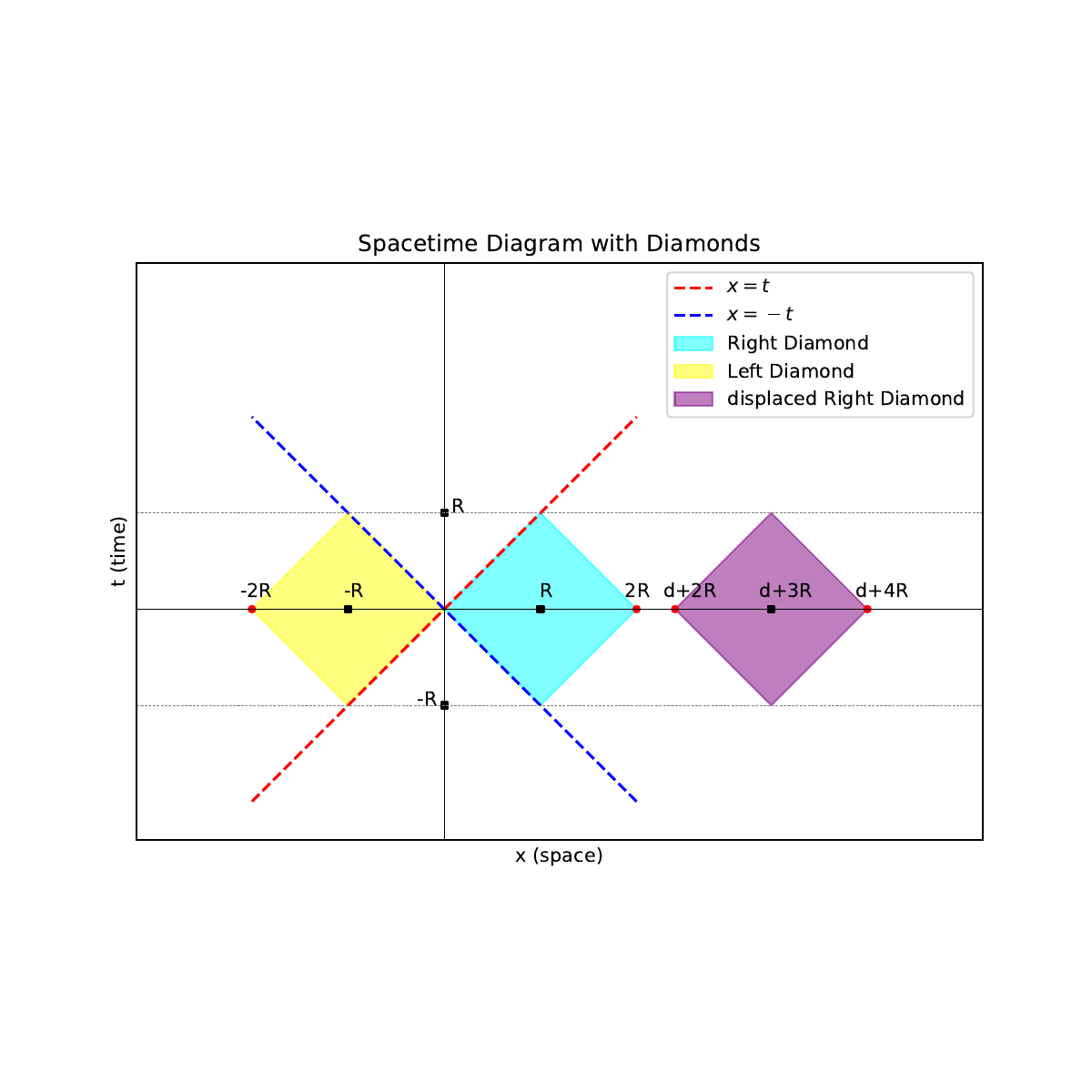}
\end{minipage} \hfill
\caption{Three diamond regions for the Mermin inequality. The distance between the purple and the blue diamonds in the right wedge is fine tuned so that the two regions are spacelike.}
\label{mmm}
\end{figure}
As before, the test functions $(g,g')$ are supported in the yellow diamond, while $(f,f')$ are supported in the blue diamond. Additionally, the pair $(h,h')$ is supported in the third purple diamond. Figure~\eqref{mmm4} represents a plot of the test functions $(f,h)$.
By construction, $(g,g')$ and $(h,h')$ are spacelike.

Concerning now the two diamonds in the right wedge, {\it i.e.}, the purple and blue ones, causality is achieved by considering values of the parameters which give rise to vanishing Pauli-Jordan expressions. In other words, only parameters for which the following conditions are fulfilled are considered for the numerical tests:
\begin{eqnarray} 
\Delta_{PJ}(h,f) & = & \int d^2x d^2y \; h(x) \;\Delta_{PJ}(x-y) \; f(y) = 0 \;, \nonumber \\
\Delta_{PJ}(h',f) & = & \int d^2x d^2y \; h'(x) \;\Delta_{PJ}(x-y) \; f(y) = 0 \;, \nonumber \\
\Delta_{PJ}(h,f') & = & \int d^2x d^2y \; h(x) \;\Delta_{PJ}(x-y) \; f'(y) = 0 \;, \nonumber \\
\Delta_{PJ}(h',f') & = & \int d^2x d^2y \; h'(x) \;\Delta_{PJ}(x-y) \; f'(y) = 0 \;, \label{PJt}
\end{eqnarray}
ensuring that causality between the blue and the purple diamonds  is fulfilled. \\\\Let us proceed with the evaluation of expression \eqref{mthree}. From Eq.~\eqref{Had}, it follows that 
\begin{eqnarray} 
\langle 0|\; A_f B_g C_h \; |0\rangle & = & 1 - 8 e^{-(H(f,f) + H(g,g) + H(f,g) + H(g,h))} + 4 e^{-(H(f,f) + H(g,g) + H(f,g))}  \nonumber \\
&+&  4 e^{-(H(g,g) + H(h,h) + H(h,g))} +  4 e^{-(H(f,f) + H(h,h) + H(f,h))} \nonumber \\
& - & 2 \left( e^{-H(f,f)} + e^{-H(g,g)} +e^{-H(h,h)} \right)  \;. \label{ABC}
\end{eqnarray}

Thus, for the correlation function \eqref{mthree}, we obtain 
\begin{eqnarray} 
\langle 0|\; {\cal M}_3 \; |0\rangle & = & 2 -8 \left(  e^{-(H(f',f') + H(g,g) + H(h,h) + H(f',g) + H(g,h))}  +  e^{-(H(f,f) + H(g',g') + H(h,h) + H(f,g') + H(g,h))}     \right)  \nonumber \\
&\; &  -  8 \left(  e^{-(H(f,f) + H(g,g) + H(h',h') + H(f,g) + H(g,h'))}  -  e^{-(H(f',f') - H(g',g') + H(h',h') + H(f',g') + H(g',h'))}     \right)  \nonumber \\
& \; & + 4 \left( e^{-( H(f',f') + H(g,g) + H(f',g) )}+ e^{-( H(g,g) + H(h,h) + H(h,g) ) } +  e^{-( H(f',f') + H(h,h) + H(f',h) )}\right)  \nonumber \\
& \; & + 4 \left(  e^{-( H(f,f) + H(g',g') + H(f,g') )}  + e^{-( H(g',g') + H(h,h) + H(g',h) )}+ e^{-( H(f,f) + H(h,h) + H(h,f) )}  \right)  \nonumber \\
& \; & + 4 \left(  e^{-( H(f,f) + H(g,g) + H(f,g) )}  + e^{-( H(g,g) + H(h',h') + H(g,h') )}+ e^{-( H(f,f) + H(f,h') + H(h',h') )}  \right)  \nonumber \\
& \; & - 4 \left(  e^{-( H(f',f') + H(g',g') + H(f',g') )}  + e^{-( H(g',g') + H(h',h') + H(g',h') )}+ e^{-( H(f',f') + H(h',h') + H(h',f') )}  \right)  \nonumber \\
& \; & - 4 \left(     e^{-H(f,f)} + e^{-H(g,g)}.  + e^{-H(h,h)}      \right)    \;. \label{exm3}
\end{eqnarray} 
For the violation of the Mermin inequality, quite good results were obtained, displayed in Table \eqref{mtab}. We can once again notice that the size of the violation increases as the mass parameter decreases, as in the Bell-CHSH case.





\begin{table}[h!]
\begin{tabular}[t]{|c|c|c|c|c|c|c|c|c|c|c|c|c|c|c|c| }
\hline
$a$ & $\eta$ & $b$ & $\sigma$ & $a'$ & $\eta'$ & $b'$ & $\sigma'$ &	$m$  & $R$ & $R'$ & $p$ & $p'$ & $\zeta$ & $\zeta'$ &
$\langle \mathcal{M}_3 \rangle$  \\
\hline
0.9465& 0.3055 & 0.1312 & 0.0749& 2.7175 & 2.4143 & 7.3920 & 9.9823 & 0.0898&
1.7299 & 2.6952 & 0.3337 & 1.2638 & 0.09370 & 0.3913 & 2.5458
\\	
\hline

0.9066& 0.2857 & 0.2634 & 0.0064 & 0.1340 & 1.6740 & 7.0886 & 0.3461 &0.0689 &
1.8967 & 2.8646 & 0.7798 & 5.1077 & 0.0462 & 0.2178  & 3.3092	\\	
\hline			
0.3106& 0.0722 & 0.1970 & 0.0334& 0.6929 & 2.1471 & 5.6812 & 6.1663 &0.0536 &
1.8416 & 2.5998 &0.6798 & 4.3208 & 0.0749 & 0.0855 & 3.3318
\\	
\hline
0.6489 & 0.0485 & 0.2419 &0.0737 & 4.5423 &  3.4910 & 4.8776& 9.7773 & 0.0339&
1.9304 & 2.6174 & 0.2551 & 0.2830 & 0.0987 & 0.0135  & 3.5607
\\	
\hline			
\end{tabular}
\caption{Results obtained for the Mermin correlation function \eqref{mthree}, corresponding to $d=2R$, $d'=2R$. The values of the violation are reported in the last column.}
\label{mtab}
\end{table}


\section{Checking the cluster property}\label{Cp}

The cluster property is one of the fundamental features of Quantum Field Theory \cite{Haag:1992hx}, which expresses the decaying behavior of the correlation functions with respect to the characteristic spatial distance of the system. Consider, for instance, the blue and purple diamonds of Fig.~\eqref{mmm}, located in the right wedge. Following \cite{Llandau,Summers:1987fn,Summers:1987squ,Summers:1987ze}, the cluster property is expressed by 
\begin{equation} 
\Big| \;\langle 0|\; A_f C_h \;|0\rangle - \langle 0|\; A_f  \;|0\rangle \;\langle 0|\; C_h \;|0\rangle \; \Big| \le e^{-md} \;, \label{cl1}
\end{equation} 
where $m$ is the mass parameter and $d$ the minimum spatial distance between the two diamonds. It is easy to notice that the cluster property can be cast into the form 
\begin{equation}
{\cal C}_{\rm cluster} \le 0 \;, \qquad {\cal C}_{\rm cluster} = e^{-(H(f,f) + H(h,h))} \Big| 1 - e^{-H(f,h)} \Big| - \frac{1}{4} e^{-md} \;, \label{cl2}
\end{equation}
which can be tested numerically. Applying our previously described random tests, the entire set of outputs for ${\cal C}_{\rm cluster}$ is negative, as required. In Table \eqref{clustt},  a sample of the values obtained for ${\cal C}_{\rm cluster}$ from our random tests is displayed.

\begin{table}[h!]
\begin{tabular}[t]{|c|c|c|c|c|c|c|c|c|c|c|c|c|c|c|c| }
\hline
${\cal C}_{\rm clust} $ & $m $ &  $d$   \\
\hline
-0.147295 & 0.145679 & 3.62183 \\
\hline
-0.0324423 & 	0.671933 & 3.02877	 \\
\hline
-0.00770447 & 0.84799 & 4.05452 \\
\hline
-0.0558507 & 0.396405 & 3.77189 \\
\hline
\end{tabular}
\caption{Check of the cluster property. The values of $m$ and $d$ are selected in a random way. According to Eq.~\eqref{cl2},  ${\cal C}_{\rm clust} $ is negative. }
\label{clustt}
\end{table}		

To conclude this section, we present the values of the remaining parameters  $(a,\eta, p, \zeta, R)$ corresponding to Table \eqref{clustertt}:

\begin{table}[h!]
\begin{tabular}[t]{|c|c|c|c|c|c|c|c|c|c|c|c|c|c|c|c| }
\hline
$ a $ & $ \eta $ & $ p $ & $d $ & $ \zeta $ & $m $ & $R $ & ${\cal C}_{\rm clust} $   \\
\hline
0.300835 
& 0.242515 
&  0.499921 
& 3.62183 &  0.677292 & 0.145679 &1.83324 &
 -0.147295 \\
\hline
0.771838 & 0.578664 &  0.709017 & 3.02877 & 0.620836 &  0.671933, & 0.896623 & 
 -0.0324423\\
\hline
0.973651 &  0.699229 & 0.670236 &  4.05452, & 0.829479, &  0.84799 &  1.46169 &
-0.00770447 \\
\hline
0.241704 & 0.010378 &  0.177163 &  3.77189 &  0.236035 &  0.396405 & 1.25455 &
 -0.0558507 \\
\hline
\end{tabular}
\caption{Values of the remaining parameters $(a,\eta, p, \zeta,R)$, as defined in Eqs.~\eqref{ff} and \eqref{hh}, corresponding to the configurations presented in Table \eqref{clustt}.}
\label{clustertt}
\end{table}





\section{Conclusion}\label{conc}


This work has demonstrated significant advances in understanding Bell-type inequalities within Quantum Field Theory through both analytical and numerical approaches. Our key contributions can be summarized as follows:

The construction of dichotomic operators using vacuum projectors and Weyl unitary operators proved highly effective for analyzing Bell-CHSH and Mermin inequalities in scalar field theory. The numerical framework developed for causal diamond regions yielded substantial violations of both inequalities, with Bell-CHSH violations approaching Tsirelson's bound of $2\sqrt{2}$ in the low-mass limit.

Particularly noteworthy is the agreement between our numerical results and the theoretical predictions of Summers and Werner regarding maximal violation in the massless limit for tangent diamonds. Despite working in 1+1 dimensional spacetime with its inherent infrared challenges, our interpolation analysis strongly supports these fundamental theorems.

The extension to three-particle Mermin inequalities represents a novel contribution, demonstrating that our framework can successfully analyze more complex quantum correlations. The observed violations, reaching values of up to 3.56, suggest rich quantum behavior in field-theoretical systems beyond the two-particle case.

Our numerical validation of the cluster property further strengthens the consistency of these results within the axioms of Quantum Field Theory. This confirmation bridges the gap between abstract algebraic requirements and concrete physical observations.

The demonstrated compatibility between numerical results and theoretical predictions suggests that our framework could serve as a valuable tool for exploring quantum correlations in field theory, potentially leading to deeper insights into the nature of quantum nonlocality in relativistic systems.

 Let us conclude by highlighting two topics currently under investigation. The first concerns the generalization of the present framework to higher-dimensional spacetimes. This extension would allow us to address the massless case from the outset, since, in contrast to the 1+1 dimensional scenario, scalar field theories in 1+2 and 1+3 dimensions are not affected by infrared singularities. However, this generalization introduces new computational challenges. For example, in 1+3 dimensions, both the Pauli–Jordan and Hadamard functions exhibit increased complexity compared to their 1+1 dimensional counterparts. Additionally, the numerical evaluation of eight-dimensional integrals becomes necessary, requiring the development or adoption of more sophisticated numerical techniques.

 The second point pertains to the potential experimental applicability of the framework we have outlined. As can be appreciated, this is a highly non-trivial matter, primarily due to the absence of a well-established measurement theory for relativistic quantum field theories, which inherently involve an infinite number of degrees of freedom. Nonetheless, it is worth noting that the operators considered in this work (see Eq.~\eqref{ABop}) are constructed from projectors onto coherent states. These coherent states are generated by the action of the Weyl operator $W_h$ on the vacuum state $ |0\rangle $, yielding $ W_h|0\rangle $. In light of this, it is conceivable to extend the present framework to the case of Maxwell theory, thereby exploring the concrete possibility of applying it to the study of entanglement in coherent electromagnetic fields---an area that has already been extensively investigated (see~\cite{Sanders:2012pme}).

\section*{Acknowledgments}
The authors would like to thank the Brazilian agencies Conselho Nacional de Desenvolvimento Científico e Tecnológico (CNPq), Coordenação de
Aperfeiçoamento de Pessoal de Nível Superior - Brasil (CAPES) and Fundação Carlos Chagas Filho de Amparo à Pesquisa do Estado do Rio de Janeiro (FAPERJ) for financial support. In particular, A. F.~Vieira is supported by a postdoctoral grant from FAPERJ in the Pós-doutorado Nota 10 program, grant No. E-
26/200.135/2025. S. P.~Sorella, I.~Roditi, and M. S.~Guimaraes are CNPq researchers under contracts 301030/2019-7, 311876/2021-8, and 309793/2023-8, respectively.


\appendix

\section{The massive real scalar field in 1+1 Minkowski spacetime}\label{appA}

The massive real scalar field in 1+1-dimensional Minkowski spacetime has the plane-wave expansion:
\begin{equation} \label{qf}
\varphi(t,x) = \int \! \frac{d k}{2 \pi} \frac{1}{2 \omega_k} \left( e^{-ik_\mu x^\mu} a_k + e^{ik_\mu x^\mu} a^{\dagger}_k \right), 
\end{equation} 
where $\omega_k  = k^0 = \sqrt{k^2 + m^2}$. For the canonical commutation relations, one has 
\begin{align}
[a_k, a^{\dagger}_q] &= 2\pi \, 2\omega_k \, \delta(k - q), \\ \nonumber 
[a_k, a_q] &= [a^{\dagger}_k, a^{\dagger}_q] = 0. 
\end{align}
It is a well-known fact that quantum fields must be considered as operator-valued distributions \cite{Haag:1992hx}. Consequently, they need to be smeared to produce well-defined operators that act on the Hilbert space, {\it i.e.}
\begin{align} 
\varphi(h) = \int \! d^2x \; \varphi(x) h(x) \;, \label{smmd}
\end{align}
where $h$ is a real smooth test function with compact support. With the smeared fields, the Lorentz-invariant inner product is introduced by means of the two-point  smeared Wightman function
\begin{align} \label{InnerProduct}
\langle f \vert g \rangle &= \langle 0 \vert \varphi(f) \varphi(g) \vert 0 \rangle =  \frac{i}{2} \Delta_{PJ}(f,g) +  H(f,g) \;, 
\end{align}
where $f$ and $g$ are also real smooth test functions with compact support, and $ \Delta_{PJ}(f,g)$ and $H(f,g)$ are the smeared versions of the Pauli-Jordan and Hadamard expressions
\begin{align}
\Delta_{PJ}(f,g) &=  \int \! d^2x d^2y f(x) \Delta_{PJ}(x-y) g(y) \;,  \nonumber \\
H(f,g) &=  \int \! d^2x d^2y f(x) H(x-y) g(y)\;. \label{mint}
\end{align}
Here, $\Delta_{PJ}(x-y)$ and $H(x-y)$ are given by
\begin{eqnarray} 
\Delta_{PJ}(t,x) & =&  -\frac{1}{2}\;{\rm sign}(t) \; \theta \left( \lambda(t,x) \right) \;J_0 \left(m\sqrt{\lambda(t,x)}\right) \;, \nonumber \\
H(t,x) & = & -\frac{1}{2}\; \theta \left(\lambda(t,x) \right )\; Y_0 \left(m\sqrt{\lambda(t,x)}\right)+ \frac{1}{\pi}\;  \theta \left(-\lambda(t,x) \right)\; K_0\left(m\sqrt{-\lambda(t,x)}\right) \;, \label{PJH}
\end{eqnarray}
where 
\begin{equation} 
\lambda(t,x) = t^2-x^2 \;, \label{ltx}
\end{equation}
and $(J_0,Y_0,K_0)$ are Bessel functions, while $m$ is the mass parameter. \\\\Both the Hadamard and Pauli-Jordan distributions are Lorentz-invariant. Notably, the Pauli-Jordan distribution, \(\Delta_{PJ}(x)\), encodes relativistic causality, as it vanishes outside the light cone. Furthermore, \(\Delta_{PJ}(x)\) and the Hadamard distribution, \(H(x)\), exhibit distinct symmetry properties: \(\Delta_{PJ}(x)\) is odd under the transformation \(x \to -x\), whereas \(H(x)\) is even. When expressed in terms of smeared fields, the commutator of the field operators takes the form 
\[
\left[\varphi(f), \varphi(g)\right] = i \Delta_{PJ}(f, g).
\]
Within this framework, causality is elegantly encapsulated by the condition $\left[\phi(f), \phi(g)\right] = 0,$
whenever the supports of $f$ and $g$ are spacelike separated.


\bibliography{refs1}

\end{document}